\documentclass{appolb}
\usepackage{graphicx}

\begin{document}
\title{Study of the $K^+K^-$ final state interaction \\ in proton -- proton and electron -- positron collisions
\thanks{Presented at the Excited QCD 2013 workshop}%
}
\author{M. Silarski
\address{Institute of Physics, Jagiellonian University, PL-30-059 Cracow, Poland}
}
\maketitle
\begin{abstract}
The strength of the kaon--antikaon interaction is a crucial quantity 
for many physics topics. It is for example, an important parameter 
in the discussion on the nature of the scalar resonances $a_{0}(980)$ and $f_{0}(980)$, 
in particular for their interpretation as a $K\bar{K}$ molecules. So far, one of the few possibilities to study this interaction
is the kaon pair production in multi particle exit channels like $pp \to ppK^+K^-$.
In this article we present the latest results of the $K^+K^-$ interaction preformed based on near threshold data gathered
at the Cooler Synchrotron COSY. We discuss also shortly perspectives for a new measurement
of the kaon--antikaon scattering length in the $e^+e^-$ collisions.
\end{abstract}
\section{Introduction}
The motivation for investigating the low energy $K^+K^-$ interaction is closely
connected with understanding of the nature of scalar resonances $f_{0}$(980) and
$a_{0}$(980). Besides the interpretation as $q\bar{q}$ mesons~\cite{Morgan},
these particles were also proposed to be $qq\bar{q}\bar{q}$ tetraquark states~\cite{Jaffe},
hybrid $q\bar{q}$/meson-meson systems~\cite{Beveren} or even quark-less
gluonic hadrons ~\cite{Johnson}.
Since both $f_{0}$(980) and $a_{0}$(980) masses are very close to the sum of the $K^{+}$ and $K^{-}$
masses, they are considered also as $K\bar{K}$ bound states~\cite{Lohse, Weinstein}. 
The strength of the $K\bar{K}$ interaction is a crucial quantity regarding the formation
of such molecules.\\
The $K^+K^-$ interaction
was studied experimentally inter alia in the $pp \to ppK^+K^-$ reaction with COSY--11
and ANKE detectors operating at the COSY synchrotron in J\"{u}lich in Germany. 
The experimental data collected systematically
below~\cite{wolke,quentmeier,winter} and above~\cite{anke,Ye,disto} the $\phi$ meson threshold revealed
a significant enhancement in the shape of the excitation function near the kinematical
threshold. On the other hand, despite the search done by the COSY--11 experiment~\cite{quentmeier,f0}
and analysis based on big data samples collected by ANKE and WASA--at--COSY experiments, there is
no clear evidence of the $K^+K^-$ pairs production through the $f_{0}$(980) or $a_{0}$(980) resonances.  
The enhancement of the excitation function near the threshold may be due to the final state interaction
(FSI) in the $ppK^+K^-$ system. Indeed, the differential spectra obtained by the COSY-11~\cite{winter,PhysRevC}
and ANKE~\cite{anke} groups indicate a strong interaction in the $pK^-$ an $ppK^-$ subsystems.
The phenomenological model proposed by the ANKE collaboration based on the factorization of the final
state interaction into interactions in the $pp$ and $pK^-$ subsystems allowed
to describe the experimental $pK^-$ an $ppK^-$ invariant mass distributions assuming an effective
$pK^-$ scattering length $a_{pK^-} = 1.5i$~fm~\cite{anke,PhysRevC}. 
However, the data very close to the kinematical threshold remain underestimated,
which indicates that in the low energy region the influence of the $K^{+}K^{-}$
final state interaction may be significant~\cite{anke,PhysRevC,wilkin}. Motivated by this 
observation the COSY--11 collaboration has estimated the scattering length of the $K^{+}K^{-}$
interaction  based for the first time on the low energy $pp \to ppK^{+}K^{-}$ Goldhaber Plot
distributions measured at excess energies of Q~=~10~MeV and 28~MeV~\cite{PhysRevC}.\\
In this article we present preliminary results of the $K^+K^-$--FSI studies
combining the Goldhaber Plot distributions established by the COSY--11
group with the experimental excitation function near threshold.
\section{Parametrization of the interaction in the $ppK^+K^-$ system}
The final state interaction model used in the presented analysis
is was based on the factorization ansatz mentioned before, with an additional
term describing the interaction of the $K^+K^-$ pair (The $pK^{+}$ 
interaction was neglected since it has to be found weak~\cite{anke}). 
We have assumed that the overall enhancement factor originating from final state
interaction can be factorized into enhancements in the proton--proton, the two $pK^-$
and the $K^+K^-$ subsystems:
\begin{equation}
F_{FSI} = F_{pp}(k_{1}) \times F_{p_{1}K^-}(k_{2}) \times F_{p_{2}K^-}(k_{3}) \times F_{K^+K^-}(k_{4})
\label{row1}
\end{equation}
where $k_{j}$ stands for the relative momentum of particles in the corresponding
subsystem.
The proton -- proton scattering amplitude was taken into account using the following
parametrization:
\[F_{pp} =
  \frac{e^{i\delta_{pp}({^{1}\mbox{\scriptsize S}_{0}})} \cdot
        \sin{\delta_{pp}({^{1}\mbox{S}_0})}}
       {C \cdot k_{1}}~,\]
where $C$ stands for the square root of the Coulomb pe\-ne\-tra\-tion factor~\cite{pp-FSI}.
The parameter $\delta_{pp}({^{1}\mbox{S}_0})$ denotes the phase shift 
calculated according to the modified Cini--Fubini--Stanghellini formula with
the Wong--Noyes Coulomb correction~\cite{noyes995,noyes465,naisse506}.
Factors describing the enhancement originating from the $pK^-$
and $K^+K^-$--FSI were instead parametrized using the scattering length approximation:
\begin{equation}
\nonumber
F_{pK^{-}}=\frac{1}{1-ika_{pK^-}}~,~~~F_{K^{+}K^{-}}=\frac{1}{1-ik_{4}~a_{K^+K^-}}~,
\label{F_ppKK}
\end{equation}
where $a_{K^+K^-}$ is the scattering length of the $K^{+}K^{-}$ interaction treated
as a free parameter in the analysis.
Since the $pK^{-}$ scattering length estimated by the ANKE group should be rather treated as
an effective parameter~\cite{anke}, in the analysis we have used more realistic $a_{pK^-}$
value estimated independently as a mean of all values summarized in Ref.~\cite{Yan:2009mr}:
$a_{pK^-} = (-0.65 + 0.78i$)~fm.\\
It has to be stressed, that within this simple model we neglect the charge -- exchange
interaction allowing for the $K^{0}K^{0}\rightleftharpoons K^{+}K^{-}$
transitions, and generating a significant cusp effect in the $K^{+}K^{-}$
invariant mass spectrum near the $K^{0}K^{0}$ threshold~\cite{dzyuba}.
However, the ANKE data can be described well without introducing the cusp
effect~\cite{dzyuba}, thus we neglect it in this analysis. We also cannot
distinguish between the isospin I~=~0 and I~=~1 states of the $K^+K^-$ system.
However, as pointed out in~\cite{dzyuba}, the production with I~=~0 is
dominant in the $pp\to ppK^+K^-$ reaction independent of the exact values
of the scattering lengths.
\begin{figure}
\centering
  \includegraphics[width=0.3\textwidth,angle=0]{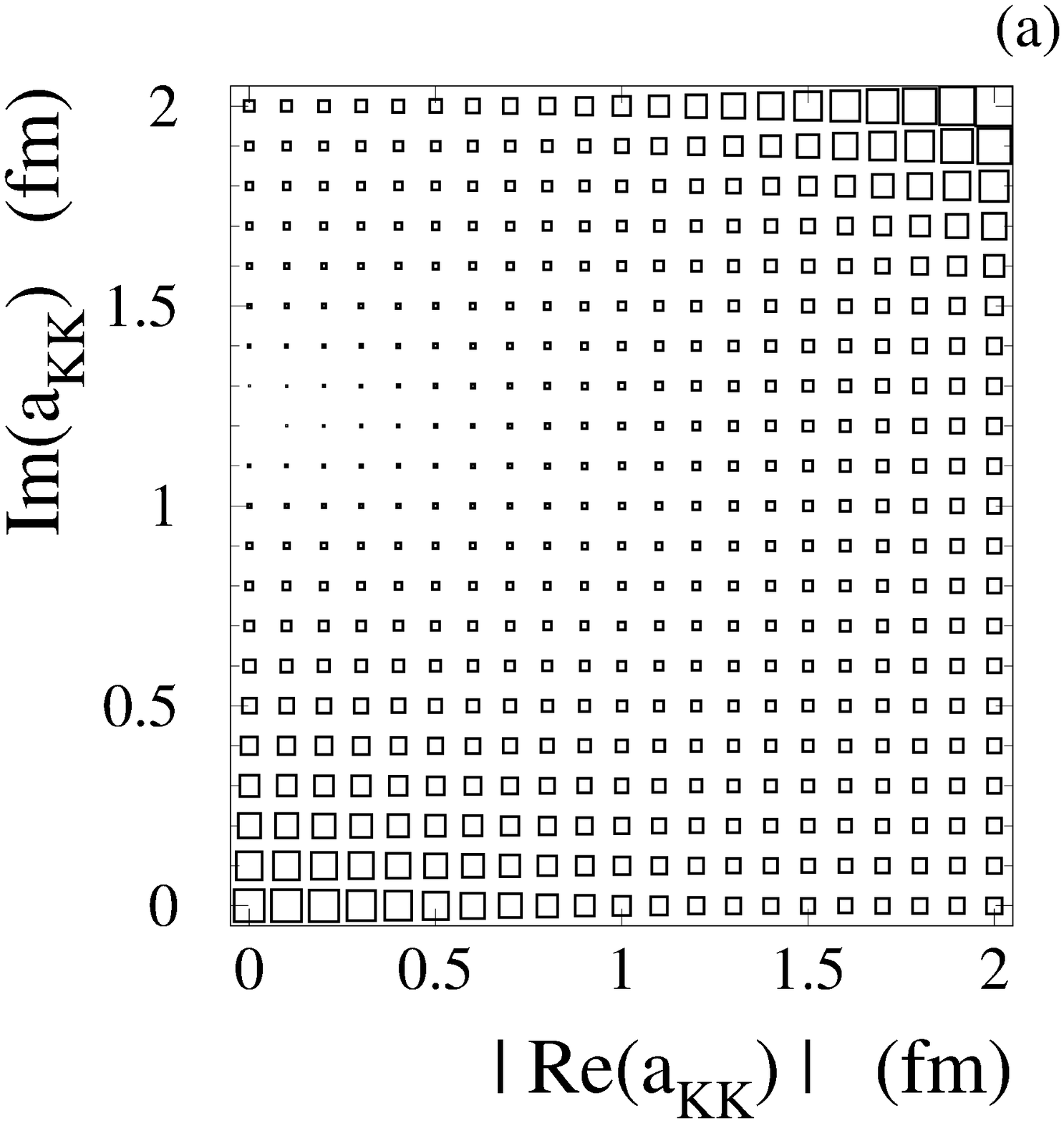}
 \includegraphics[width=0.3\textwidth,angle=0]{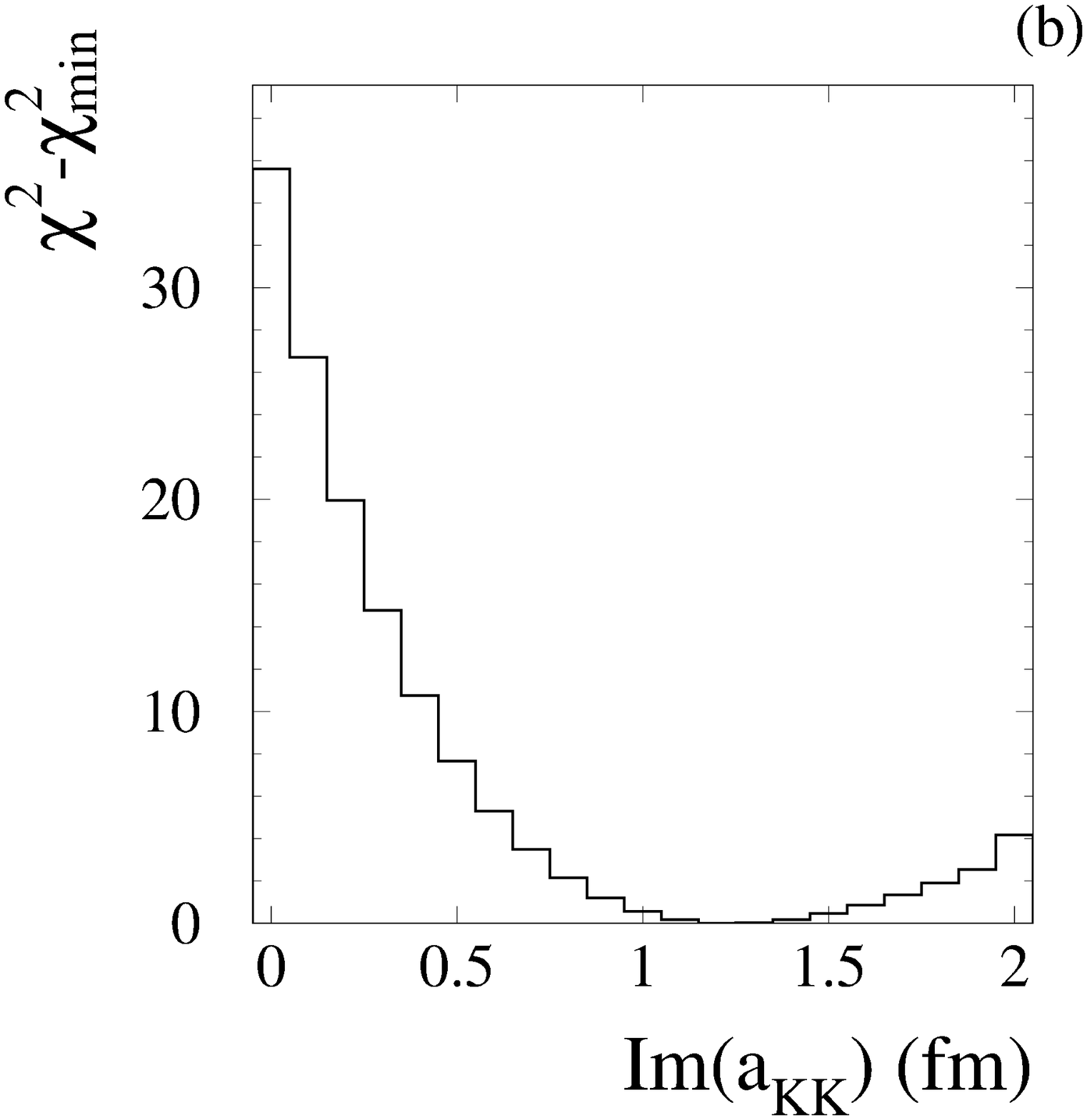}
  \includegraphics[width=0.3\textwidth,angle=0]{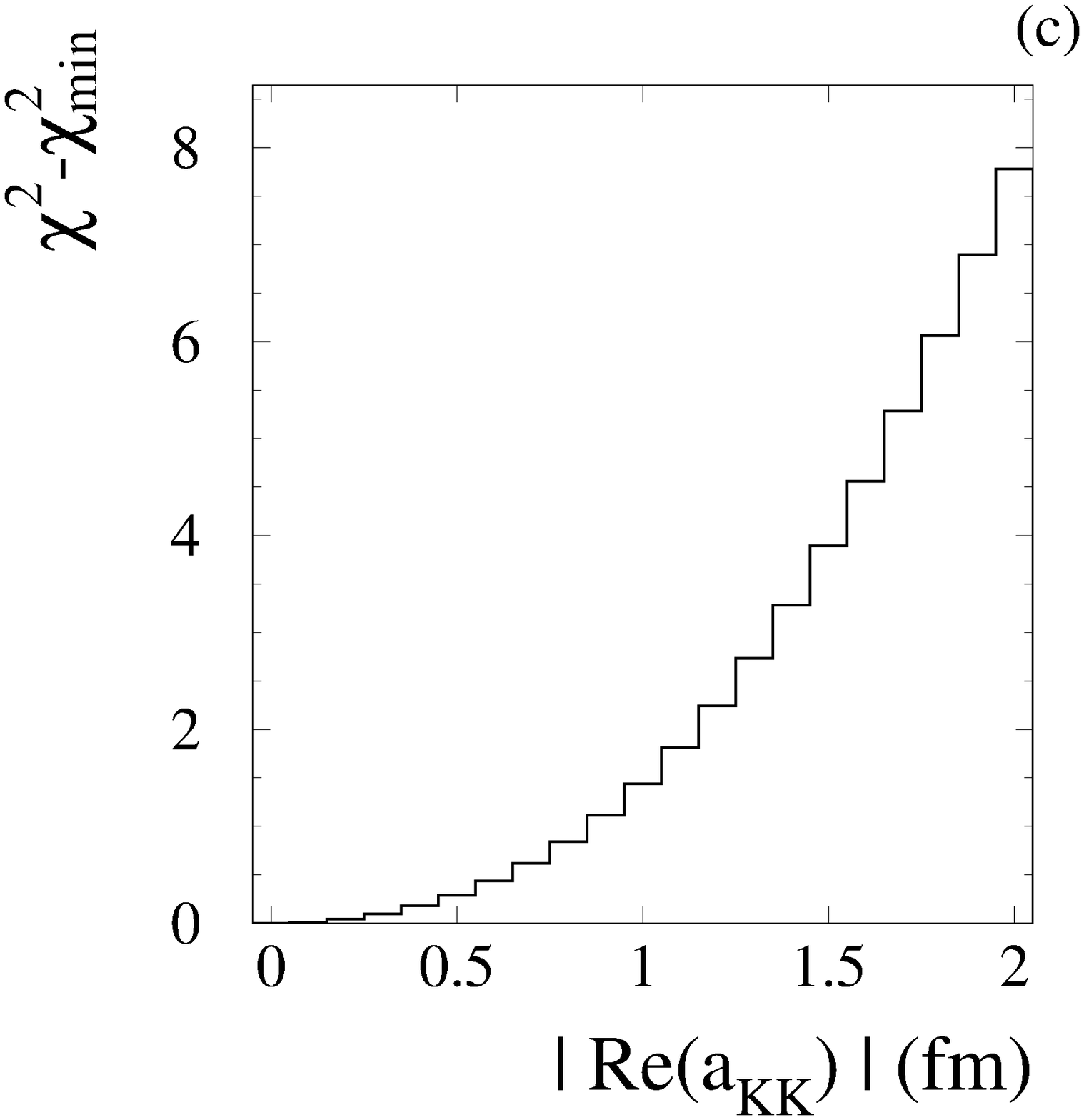}
\caption{$\chi^{2}$~-~$\chi^{2}_{min}$ distribution as a function of
$|\mathrm{Re}(a_{K^{+}K^{-}})|$ and $\mathrm{Im}(a_{K^{+}K^{-}})$(left), 
as well as its projections on each axis (center and right). $\chi^{2}_{min}$ denotes
the absolute minimum with respect to parameters $\alpha$, $|\mathrm{Re}(a_{K^{+}K^{-}})|$,
and $Im(a_{K^+K^-})$.
In the figure on the left the area of the squares is proportional
to the $\chi^{2}$~-~$\chi^{2}_{min}$ value.}
\label{fig:2}
\end{figure}
\section{Fit to the experimental data}
\begin{figure}
\centering
\includegraphics[width=0.4\textwidth,angle=0]{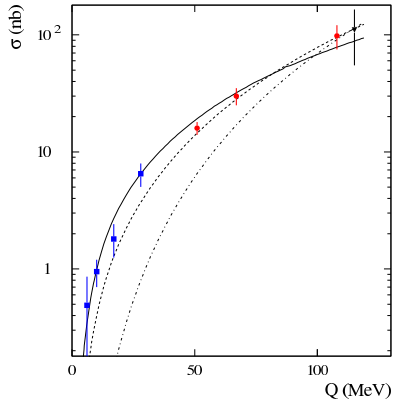}
\caption{Excitation function for the $pp\rightarrow ppK^{+}K^{-}$ reaction.
Triangle and circles represent the DISTO and ANKE measurements, respectively~\cite{anke,disto}.
The squares are results of the COSY--11~\cite{wolke,quentmeier,PhysRevC} measurements.
The dashed--dotted, dotted and solid curves represent the energy dependence
obtained assuming that there is no interaction between particles,
assuming the $pp$ and $pK^-$~--~FSI and taking into account $pp$, $pK$ and $K^+K^-$ interaction, respectively.
The dashed and dashed -- dotted curves are normalized to the DISTO data point at Q~=~114~MeV. }
\label{fig:3}
\end{figure}

In order to estimate the strength of the $K^+K^-$ interaction the experimental Goldhaber
plots, determined at excess energies of Q~=~10~MeV and Q~=~28~MeV~\cite{PhysRevC},
were compared together with the total cross sections to the results of the Monte Carlo
simulations treating the $K^+K^-$ scattering length $a_{K^{+}K^{-}}$ as an unknown parameter.
We have constructed the following $\chi^{2}$ statistics:
\begin{equation}
\nonumber
\chi^2\left(a_{K^+K^-},\alpha\right) = \sum_{i=1}^{8}\frac{\left(\sigma_{i}^{exp}
- \alpha\sigma_{i}^{m}\right)^2} {\left(\Delta\sigma_{i}^{exp}\right)^2}\\
+2 \cdot \sum_{j=1}^{2}\sum_{k=1}^{10} \, [\beta_{j} N_{jk}^s - N_{jk}^e +  N_{jk}^e \,
{\mathrm{ln}}(\frac{N_{jk}^e}{\beta_{j} N_{jk}^s})] ,
\label{eqchi2_mh}
\end{equation}
where the first term was defined following the Neyman's $\chi^{2}$ statistics,
and accounts for the excitation function near threshold for the $pp \to ppK^{+}K^{-}$
reaction. $\sigma_{i}^{exp}$ denotes the $i$th experimental total cross section
measured with uncertainty $\Delta\sigma_{i}^{exp}$ and $\sigma_{i}^{m}$
stands for the calculated total cross section normalized with a factor $\alpha$
treated as an additional parameter of the fit. 
$\sigma_{i}^{m}$ was calculated for each excess energy $Q$ as a phase space integral
over five independent invariant masses~\cite{nyborg}.
The second term of Eq.~\ref{eqchi2_mh} corresponds to the Poisson likelihood
$\chi^2$~\cite{baker} describing the fit to the Goldhaber plots determined at excess energies
$Q = 10$~MeV ($j$~=~1) and $Q = 28$~MeV ($j$~=~2) using COSY--11 data~\cite{PhysRevC}.
$N_{jk}^e$ denotes the number of events in the $k^{th}$ bin of the $j^{th}$ experimental
Goldhaber plot, and $N_{jk}^s$ stands for the content of the same bin in the simulated
distributions. The simulations were normalized with a factor defined for the $j^{th}$
excess energy as: $\beta_{j} = \frac{L_{j}\alpha\sigma_{j}^{m}}{N_{j}^{gen}}$.
Here $L_{j}$ stands for the total luminosity ~\cite{winter} and $N_{j}^{gen}$ denotes
the the total number of simulated $pp \to ppK^+K^-$ events.
The $\chi^2$ distribution obtained after subtraction of its minimum value is presented
in Fig.~\ref{fig:2} as a function of the real and imaginary part of the $K^+K^-$ scattering
length. The best fit to the experimental data corresponds to:
\begin{eqnarray}
\nonumber
\left|\mathrm{Re}(a_{K^{+}K^{-}})\right| = 0.0^{~+1.1_{stat}}_{~-0.0_{stat}}~\mathrm{fm},~~~~
\mathrm{Im}(a_{K^{+}K^{-}}) = 1.1^{~+0.6_{stat}~+0.9_{sys}}_{~-0.5_{stat}~-0.6_{sys}}~\mathrm{fm},
\label{chi2results}
\end{eqnarray}
with a $\chi^2$ per degree of freedom of: $\chi^2/ndof = 1.87$.
The statistical uncertainties were determined at the 70\% Confidence Level (C.L.)
taking into account that in the case of the three fit parameters~\cite{james}.
Systematic errors due to the uncertainties
in the assumed $pK^{-}$ scattering length were instead estimated as a maximal difference
between the obtained result and the $K^+K^{-}$ scattering length determined using different
$a_{pK^{-}}$ values quoted in Ref.~\cite{Yan:2009mr,Martin}.
In case of the $\left|\mathrm{Re}(a_{K^{+}K^{-}})\right|$ the differences were negligible.\\
The final state interaction enhancement factor $|F_{K^{+}K^{-}}|^{2}$
in the scattering length approximation is symmetrical
with respect to the sign of $\mathrm{Re}(a_{K^{+}K^{-}})$, therefore he have determined only
its absolute value.
The result of the analysis with inclusion of the interaction in the $K^+K^-$ system described
in this article is shown as the solid curve in Fig.~\ref{fig:3}.
One can see that it describes quite well the experimental data over the whole energy range.
\section{Summary and outlook}
\label{secEnd}
A combined analysis of both total and differential cross section
distributions for the $pp\rightarrow ppK^{+}K^{-}$
reaction in the framework of a simple factorization ansatz allowed to
estimate the $K^+K^-$ scattering length by a factor five more
precise than the previous one~\cite{PhysRevC}. However, the determined
$a_{K^{+}K^{-}}$ value is still consistent with zero, which indicates
that in the $ppK^+K^-$ system the interaction between protons and
the $K^-$ meson is dominant. 
All studies of the $pp\rightarrow ppK^{+}K^{-}$ reaction suggest also
that the resonant $K^{+}K^{-}$ pair production near threshold proceeds
rather through the $Lambda(1405)$ resonance than through scalar
$a_0(980)$/$f_0(980)$ mesons~\cite{anke}.\\
Therefore, precise determination of the kaon--antikaon scattering
length requires less complicated final states like $K^+K^-\gamma$,
where only kaons interact strongly.
This final state can be studied for example via the $e^+e^- \to K^+K^-\gamma$
reactions with the KLOE--2 detector operating at the DA$\Phi$NE
$\phi$--factory~\cite{AmelinoCamelia}. Analysis of the invariant
mass distributions obtained in this reaction would allow detailed studies
of the $K^+K^-$--FSI, including the contribution from the production through
scalar resonances. Thus, it would be a continuation of the $a_0(980)$
and $f_0(980)$ studies done so far by the KLOE collaboration~\cite{KKg,etapi0g,pi0pi0,pi+pi-}.
\section*{Acknowledgments}
This work was supported by the Polish National Science Center
through the Grants No. 0469/B/H03/2009/37, 0309/B/H03/2011/40, 
DEC-2011/03\linebreak/N/ST2/02641, 2011/01/D/ST2/00748, 2011/03/N/ST2/02652,
2011/03/\linebreak N/ST2/02641 and by the Foundation for Polish Science through
the MPD programme and the project HOMING PLUS BIS/2011-4/3.
  

\begin{thebibliography}{9}
\bibitem{Morgan}
D.~Morgan,~M.~R.~Pennington,~Phys. Rev.~D~\textbf{48}, 1185 (1993).
\bibitem{Jaffe}
R.~L.~Jaffe,~Phys. Rev.~D \textbf{15}, 267 (1977).
\bibitem{Beveren}
E.~Van Beveren,~\textit{et al.}, Z. Phys.~C~\textbf{30}, 615 (1986).
\bibitem{Johnson}
R.~L.~Jaffe,~K.~Johnson,~Phys. Lett. \textbf{B60}, 201 (1976).
\bibitem{Lohse}
D.~Lohse~\textit{et al.},~Nucl. Phys. \textbf{A516}, 513 (1990).
\bibitem{Weinstein}
J.~D.~Weinstein,~N.~Isgur,~Phys. Rev.~D~\textbf{41}, 2236 (1990).
\bibitem{wolke}
M.~Wolke,~PhD thesis, IKP J{\"u}l-3532 (1997).
\bibitem{quentmeier}
  C.~Quentmeier~\textit{et al.},~Phys. Lett. \textbf{B515}, 276 (2001).
\bibitem{winter}
  P.~Winter~\textit{et al.},~Phys. Lett. \textbf{B635}, 23 (2006).
\bibitem{anke}
Y.~Maeda {\it et al.}, Phys.\ Rev.\ C {\bf 77}, 015204 (2008).
\bibitem{Ye}
  Q.~J.~Ye~\textit{et al.},
  Phys.\ Rev.\ C \textbf{85}, 035211 (2012).
\bibitem{disto}
F.~Balestra~\textit{et al.},~Phys. Lett. \textbf{B468}, 7 (1999).
\bibitem{f0}
P.~Moskal~\textit{et al.}, J. Phys. G \textbf{29}, 2235 (2003).
\bibitem{PhysRevC}
M.~Silarski~\textit{et al.},~Phys. Rev.~C \textbf{80}, 045202 (2009).
\bibitem{wilkin}
C. Wilkin, Acta Phys. Polon. Supp. \textbf{2}, 89 (2009).
\bibitem{pp-FSI}
P.~Moskal {\it et al.}, Phys.\ Lett.\ {\bf B482}, 356 (2000).
\bibitem{noyes995}
H.~P.~Noyes, H.~M.~Lipinski, Phys.\ Rev.\ C {\bf 4}, 995 (1971).
\bibitem{noyes465}
  H.~P.~Noyes, Ann.\ Rev.\ Nucl.\ Part.\ Sci.\  {\bf 22}, 465 (1972).
\bibitem{naisse506}
J.~P.~Naisse,~Nucl. Phys. A \textbf{278}, 506 (1977).
\bibitem{Martin}  
A. D. Martin, Nucl. Phys. \textbf{B179}, 33 (1981).
\bibitem{Yan:2009mr}
  Y.~Yan,
  arXiv:0905.4818 [nucl-th].
\bibitem{dzyuba}
A. Dzyuba~\textit{et al.}, Phys. Lett. \textbf{B668}, 315 (2008).
\bibitem{nyborg}
P.~Nyborg~\textit{et al.},~Phys. Rev. \textbf{140}, 914 (1965).
\bibitem{baker}
S.~Baker and R.~D.~Cousins, Nucl.\ Instrum.\ Meth.\  {\bf 221}, 437 (1984).
\bibitem{james}
F.~James, Comput. Phys. Commun. \textbf{20}, 29 (1980). 
\bibitem{AmelinoCamelia}
  G.~Amelino-Camelia {\it et al.},
  Eur.\ Phys.\ J.\ C \textbf{68}, 619 (2010).
\bibitem{KKg}
  F.~Ambrosino {\it et al.},
  Phys.\ Lett.\ B {\bf 679}, 10 (2009).
\bibitem{etapi0g}
  F.~Ambrosino {\it et al.},
  Phys.\ Lett.\ B {\bf 681}, 5 (2009).
\bibitem{pi0pi0}
  F.~Ambrosino {\it et al.},
  Eur.\ Phys.\ J.\ C {\bf 49}, 473 (2007). 
\bibitem{pi+pi-}
  F.~Ambrosino {\it et al.},
  Phys.\ Lett.\ B {\bf 634}, 148 (2006). 
\end{thebibliography}
\end{document}